\documentclass[cup5b]{cupbook}
\usepackage{latexsym}
\usepackage{amssymb}
\usepackage{amsfonts}
\usepackage{graphicx}
\usepackage[usenames]{color}

\begin{document}

\author[K. Bardakci and M.B.Halpern]{K. Bardakci and
 M.B.Halpern\\
 Department of Physics, University of California,\\
Theoretical Physics Group, Lawrence Berkeley National Laboratory \\
University of California, Berkeley CA 94720 USA}

\chapter[The Dual Quark Models]{The Dual Quark Models \footnote{To appear
in the volume ``The Birth of String Theory'', Ed. P. Di Vecchia (Cambridge
University Press, Cambridge UK)}}

\begin{abstract}
We briefly recall the historical environment around our 1971 and 1975
constructions of current-algebraic internal symmetry on the open string. These
constructions included the introduction of world-sheet fermions, the
independent discovery of affine Lie algebra in physics (level one of affine
$su(3)$), the first examples of the affine-Sugawara and coset constructions, and
finally -- from compactified spatial dimensions on the string -- the first
vertex-operator constructions of the fermions and level one of affine $su(n)$.
\end{abstract}

\section{Introduction}

In this contribution, we describe the environment around our 1971 paper ``New
Dual Quark Models'' \cite{BHBardakciHalpern71} and the two companion papers
``The Two Faces of a Dual Pion-Quark Model'' \cite{BHHalpern71c} and ``Quantum
Solitons which are SU(N) Fermions'' \cite{BHHalpern75} in 1971 and 1975, which
laid the foundations of non-Abelian current-algebraic internal
symmetry on the string.

The  background for our contributions included many helpful discussions 
with other early workers, including H.M.Chan, C. Lovelace, H. Ruegg 
and C. Schmidt (during KB's 1969 visit to CERN)
as well as R. Brower, S. Klein, C. Thorn, M. Virasoro and J. Weis (with MBH 
at Berkeley). Both of us also acknowledge many discussions with Y. 
Frishman, G. Segre, J. Shapiro and especially S. Mandelstam. Later 
discussions with I. Bars and W. Siegel are also acknowledged by MBH.
  With apologies to many other authors then, we reference here only the work
which was most influential in our early thinking: before Veneziano, there had
been widespread interest in the quark model of Gell-Mann and Zweig, including
the four-dimensional current algebra (Adler and Dashen \cite{BHAdlerDashen68})
of quarks.


\section{The Bosonic Dual Model}

In 1970, the work of Fubini and Veneziano \cite{BHFubiniVeneziano69} together
with the Virasoro algebra \cite{BHVirasoro70} provided an {\it algebraic
formulation} of open bosonic string theory which we found quite exciting.
The first ingredients of this formulation are the bosonic oscillators 
\begin{eqnarray}
&&[J_{\mu}(m), J_{\nu}(n)]= -m\,g_{\mu\,\nu}\,\delta_{m+n,0},\nonumber\\
&&m, n\in {\mathbb Z},\,\,
g= {\rm diag}(1,-1,\dots,-1)\ ,
\qquad
\mu,\nu= 0,1,... 25
\end{eqnarray}
which we have rescaled here as the modes of an Abelian current 
algebra on the circle. The conformal-field-theoretic convention 
for the local currents associated to these modes are
\begin{equation}
J_{\mu}(z)=\sum_{m\in {\mathbb Z}}\,J_{\mu}(m)\,z^{-m-1} =i\,\partial_{z}
Q_{\mu}(z)
\end{equation}
where 
\begin{equation}
Q(z) = q -iJ(0)ln(z) + i\sum_{m\neq 0} J(m) \frac{z^{-m}}{m}\ 
\end{equation}
are the Fubini-Veneziano fields and $\{q\}$ is canonically conjugate to the
momenta $\{p=J(0)\}$.  Normal-ordered exponentials of the Fubini-Veneziano
fields are the string vertex operators, to which we shall return below.

In our discussions we shall emphasize the 
mode algebras, but for various modern applications the 
complex variable $z$ in the local fields can be taken in many forms, including
 $exp (i\theta)$ (the circle or loop), $exp (i(\tau 
+\sigma))$ (left-movers on a closed string), the Euclidean version
$exp (\tau +i\sigma)$ for conformal field theory (CFT), or the real line for 
open strings. These alternate forms led to some confusion in 
the early days of string theory, but in fact most of 
the early work reviewed in this article was in the context of the 
open string, using the parametrization $z =exp (i\theta)$ on the circle.

Finally, the Virasoro algebra of the critical open bosonic string is
\begin{eqnarray}
L(m)&=&-\frac{1}{2}\,g^{\mu\,\nu}\,\sum_{n\in {\mathbb Z}} :J_{\mu}(n)\,
J_{\nu}(m-n):\cr
[L(m),L(n)]&=&(m-n)\, L(m+n)+\frac{26}{12}\, m\, (m^{2} -1)\,
\delta_{m+n, 0}
\end{eqnarray}
where the symbol $:\ldots:$ is bosonic normal-ordering. Under these 
generators the Abelian currents $\{J(z)\}$ covariantly transform 
with conformal weight equal to one.

Taken together, the algebraic statements in Eqs. (1.1)-(1.3) provided the
first example of modern CFT (see Belavin,Polyakov and Zamolodchikov
\cite{BHBPZ}).  We have here taken an historical liberty,
because the central extension of the Virasoro algebra was not observed until
1974, and even then only in a private communication from J. Weis quoted in
Ref. \cite{BHChodosThorn74} (Chodos and Thorn).  Throughout this article we
adopt CFT conventions, including an extra factor $z^{-\Delta}$ (see Eq. (1.1))
for the mode expansions of Virasoro primary fields with conformal weight
$\Delta$. The CFT convention guarantees that a given Virasoro primary field
creates a Virasoro primary state of the same conformal weight as $z$ goes to
$0$ on the vacuum (in this case the zero-momentum vacuum $|\{
p=J(0)=0\}\rangle$ of the open bosonic string).  On the other hand, the extra
factor $z^{-\Delta}$ must be removed for exact correspondence of the local
fields with $z =exp (i\theta)$ in early work on the open string.


\section{World-Sheet Fermions}

As it appeared then, the open bosonic string had two deficiencies beyond 
the tachyonic ground state. In the 
first place Lie-algebraic internal symmetries were implemented only by the
multiplicative Chan-Paton factors \cite{BHPatonChan69}, which live on 
the boundary of the string. Second, 
the bosonic string had no spin, that is, no space-time fermions. Our
first attempt to solve the second problem was a generalization of 
Chan-Paton factors to include spin \cite{BHBardakciHalpern69}, 
but this multiplicative approach was unsuccessful
because it included negative-norm states. 

It seemed natural then to approach the problems of spin and internal 
symmetry in a different way, by introducing new local ``quark'' fields on the 
string -- now called {\it world-sheet fermions} \cite{BHBardakciHalpern71}:
\begin{equation}
\bar{\psi}^{I}(z)\,=\,\sum_{p\in {\mathbb Z}+\frac{1}{2}}\bar{\psi}^{I}(p)
z^{-p-\frac{1}{2}}, \,\,\,\,\,
\psi_{I}(z)\,=\,\sum_{p\in {\mathbb Z}+\frac{1}{2}}\psi_{I}(p)
z^{-p-\frac{1}{2}}
\end{equation}
whose modes obey anticommutation relations
\begin{eqnarray}
[\psi_{I}(q), \psi_{J}(p)]_{+}&=&[\bar{\psi}^{I}(q),\bar{\psi}^{J}(p)]_{+}
=0,\cr
[\psi_{I}(q),\bar{\psi}^{J}(p)]_{+}&=&\delta^{J}_{I}\,\delta_{q+p,0},
\,\,\,\,\,\,q, p\in {\mathbb Z}+\frac{1}{2}.
\end{eqnarray}
 As in the four-dimensional quark model itself, the generic indices 
$\{I,J\}$ were designed so that the world-sheet fermions could be 
 space-time spinors $\{\psi_{r}\}$ or  carry internal symmetry or both 
$\{\bar\psi \gamma_{\mu}\lambda_{a}\psi\}$. 
The Virasoro generators of our world-sheet fermions are
\begin{equation}
L(m)=\sum_{p\in {\mathbb Z}+\frac{1}{2}}\left(p+\frac{m}{2}\right)\,
:\bar{\psi}^{I}(p)\,
\psi_{I}(m -p): 
\end{equation}
with fermionic normal-ordering and central charge $c$ equal to the number 
of ($\psi, \bar\psi$) pairs. In what follows, 
we often refer to our world-sheet fermions as the {\it dual quarks}.  

Under these Virasoro generators, the dual quark fields $\{\psi,\bar{\psi}\}$
are Virasoro primary fields with conformal weight $1/2$ and trivial monodromy.
The half-integer moding of the dual quark fields guaranteed a unique vacuum
(or ground state )$|0\rangle$ which is annihilated by the positive modes of
both fields, and the adjoint operation is defined for internal symmetry as $
{\psi (p)}^{\dagger} = \bar{\psi} (-p)$ (with a $\gamma_{0}$ modification for
spin). In either case, our Virasoro generators satisfy the generalized
hermiticity ${L(m)}^{\dagger} = L(-m)$ and the vacuum state conditions:
\begin{equation}
L(m)|0\rangle =0 \ ,\qquad m \ge -1\ .
\end{equation}
  
With internal symmetry only, our dual quark models then provided the first
examples of {\it unitary} CFT, with a positive-definite Hilbert space on a
non-degenerate vacuum with conformal weight $\Delta = 0$.  In modern parlance,
the half-integer moded complex BH fermions are related to anti-periodic Weyl
fermions on the world sheet, but it should be born in mind that fermionic
world-sheet action formulations of open string theory had not been studied at
that time.

Were these the first world-sheet fermions? In fact, our work was essentially
simultaneous with that of Ramond \cite{BHRamond71}: 
Our paper was submitted in November of 1970 while Ramond's paper with
integer-moded Majorana-Weyl world-sheet fermions was
submitted in January 1971; the two papers were
published in the same issue of Physical Review. As we now know the R fermions
$\{b_{\mu}\}$, with their somewhat-counterintuitive space-time vector label
$\mu$ and superconformal symmetry, provided the correct covariant description
of space-time fermions and -- in a perfect match to the R construction --
half-integer moded Majorana-Weyl world-sheet fermions $\{b_{\mu}\}$ were later
introduced for the bosonic sector by Neveu and Schwarz
\cite{BHNeveuSchwarz71}.

For the historical record, the lure of our seemingly more physical space-time
spinor fields $\{\psi_{r}\}$ persisted for some time, leading to many
connections (Halpern \cite{BHHalpern71a}, \cite{BHHalpern71b},
\cite{BHHalpern71c}, Bardakci \cite{BHBardakci71}, Halpern and Thorn
\cite{BHHalpernThorn71a}, \cite{BHHalpernThorn71b} and Mandelstam
\cite{BHMandelstam73a}, \cite{BHMandelstam73b}) between our fields 
and those of the RNS formulation (We mention in
particular the first example of half-integer-moded scalar fields in
\cite{BHHalpernThorn71b}). 
Our original intuition was however not realized until 1982, when
a world-sheet fermion with 10-dimensional Majorana-Weyl spinor indices was
successfully incorporated by Green and Schwarz in the light-cone formulation
of superstring theory \cite{BHGreenSchwarz82}.

In what follows, we therefore restrict the discussion primarily to our
various current-algebraic and CFT constructions on the open string.

\section{Affine Lie algebra}

For continuity with the bosonic string, we begin this discussion with
a footnote of our 1971 paper -- which observed that the spatial components 
$\{I=1,\ldots,25\}$ of
the Abelian currents (1.1),(1.2) could be constructed
out of our world-sheet fermions \cite{BHBardakciHalpern71}: 
\begin{eqnarray}
J_{I}(m)&=& \sum_{p\in {\mathbb Z}+\frac{1}{2}}:\bar{\psi}^{I}(p)\,
\psi_{I}(m- p):,\cr
J_{I}(z)&=&:\bar{\psi}^{I}(z)\,\psi_{I}(z):\,\,\,=\,\,\,i\,\partial_{z} Q_{I}(z).
\end{eqnarray}
This result is now recognized as a simple but important part of 
Bose-Fermi equivalence on the open string, a subject to which we will 
return below.

Our next construction was the independent discovery of affine Lie 
algebra in physics \cite{BHBardakciHalpern71}, still sometimes 
called the ``Dual Quark Model'', in which
the non-abelian currents are naturally constructed as dual-quark 
bilinears with conformal weight 1:
\begin{equation}
J_{a}(z)\,=\,\sum_{m\in \mathbb Z} J_{a}(m) z^{-m-1} \,=\, 
: \bar\psi(z)T_{a}\psi(z):
\end{equation}

\begin{eqnarray}
J_{a}(m)&=&\sum_{p\in {\mathbb Z}+\frac{1}{2}} :\bar{\psi}^{I}(p)\,
{\left(T_{a}\right)_{I}}^{J}\,\psi_{J}(m- p):,\cr
[T_{a}, T_{b}]&=&i\,{f_{a b}}^{c}\, T_{c},\,\,\,\,\,\,
 Tr\left(T_{a}\,T_{b}\right)
=k\,\eta_{a b}.
\end{eqnarray}
Here $\{T\}$ can be any irreducible matrix
representation of any  simple Lie algebra $g$, with structure constants 
$\{f\}$ and  Killing metric 
$\{\eta\}$. Then
one finds that the modes of the dual quark-model currents $\{:\bar\psi T 
\psi:\}$ are the generators of the
{\it affine Lie algebra}
\begin{eqnarray}
[J_{a}(m), J_{b}(n)]&=&i\,{f_{a b}}^{c}\,J_{c}(m+n)+ m\,\left(k\,
\eta_{a b}\right)\,\delta_{m+n, 0},\cr
a,\,b,\,c &=&1 . . .dim(g),\,\,\,\,\,\,\,\,m,\,n\in {\mathbb Z}
\end{eqnarray}
where the quantity $k$ in the central term is called the {\it level} of the
affine algebra. (We assume in this article that the highest root-length
squared of Lie $g$ is two, in which case the level in the dual quark models is
also equal to the Dynkin index of irreducible representation T.) Affine Lie algebra is also known
as centrally- extended non-Abelian current algebra on the circle.
In this construction, the dual-quark fields  $\{\psi, \bar{\psi}\}$ transform 
respectively in the $\{T,\bar{T}\}$ representations of Lie $g$ -- the Lie 
algebra being generated by the zero modes $\{J_{a}(0)\}$ of the affine algebra.

In fact our first example of this construction used the $3\times3$ 
Cartesian Gell-Mann 
matrices $\{T_{a}\propto \lambda_{a}\}$ of $su(3)$ (and the Dynkin 
index of the fundamental of $su(n)$ is 1) -- so the \emph {first concrete
representation of affine Lie algebra} was $su(3)_{1}$, that is, level one of
affine $su(3)$ (see also Sec. 1.8). In this case, as in the
four-dimensional quark model itself, the dual-quark fields
$\{\psi,\bar{\psi}\}$ transform respectively in the $(3,\bar3)$
representations of $su(3)$.  Dual quark-model constructions of the currents of
$so(3,1)_{1}$ and $ so(4,1)_{1}$ were also included in our 1971 paper, and
further studied along with those of higher Lorentz algebras in the first
companion paper \cite{BHHalpern71c}.  The dual quark models of $su(n)_{1}$
were given later in the second companion paper \cite{BHHalpern75}. 

Because connections between open strings and field theory were not
understood at the time, our interpretation of the central terms in 
our string-theoretic constructions 
of affine Lie algebras on the circle was based on the 
presence of such a term in the Abelian mode algebra (1.1) of the 
open bosonic string -- to which our constructions reduced in the 
Abelian case. This understanding was supported by our fermionic 
construction (see Eq. (1.7) above) of the modes of the Abelian 
current algebra.

On the other hand, we often think now of the central term in affine Lie
algebra as an analogue on the circle of the equal-time Schwinger term
\cite{BHSchwinger59} (indeed, see Eq. (3.10) in
Ref. \cite{BHBardakciHalpern71}), so we briefly sketch the history of $su(n)$
current algebra in the two-dimensional Lorentz-invariant quantum field theory
of massless free fermions: The equal-time Schwinger term in $su(n)$ current
algebra on the line was first computed in 1969 (see Coleman, Gross and Jackiw
\cite{BHColemanGrossJackiw69}), and the one-dimensional $su(n)$ current
algebras of the light-cone currents $\{ J_{a}^{\pm}(t \pm x)\}$ -- with
one-dimensional Schwinger terms -- were given later by Dashen and
Frishman \cite{BHDashenFrishman73}. In fact however the momenta are
continuous in this non-compactified, Lorentz-invariant context, so our
open-string mode forms of the fermions and affine Lie algebra on the circle
\cite{BHBardakciHalpern71} do not appear in these papers.

For a modern introduction to affine Lie algebras in CFT, see the work
by Gepner and Witten \cite{BHGW} and also the representation theory for all 
integer levels in  Kac \cite{BHKac90}. In physics the affine algebras
have sometimes also been called Kac-Moody algebras, but this term is more properly
reserved for the general algebraic systems including affine and hyperbolic 
algebras (see Sec. 1.8).

\section{ The Affine-Sugawara Constructions}

The {\it affine-Sugawara constructions} \cite{BHBardakciHalpern71} \cite{BHHalpern71c} are the simplest realization of  
Virasoro generators as quadratic forms on the generators of affine Lie algebras ,
\begin{eqnarray}
L_{g}(m)&=&\frac{\eta^{a b}}{2\,k+Q_{g}}\,\sum_{n\in {\mathbb Z}} :J_{a}(n)\,
J_{b}(m - n):,\cr
[L_{g}(m), L_{g}(n)]&=&(m- n)\,L_{g}(m+n)+\frac{c_{g}}{12}\,m\,
(m^{2} -1)\,\delta_{m+n,0},\cr
c_{g}&=&\frac{2\,k\,dim(g)}{2\,k+Q_{g}}.
\label{suga}
\end{eqnarray}
The quantity $Q_{g}$ is the quadratic Casimir of Lie $g$, which in our present
root-normalization is twice the dual Coxeter number of $g$.  The original
Sugawara model \cite{BHSugawara68}, Sommerfield \cite{BHSommerfield68},
Bardakci, Frishman and Halpern \cite{BHBardakciFrishmanHalpern68}, Bardakci
and Halpern \cite{BHBardakciHalpern68}, Gross and Halpern
\cite{BHGrossHalpern69}, Coleman, Gross and Jackiw
\cite{BHColemanGrossJackiw69} was in four dimensions on a different current
algebra, the so-called algebra of fields.  The 
affine-Sugawara (or Sugawara-like) constructions in Eq. (1.11) hold for all integer levels
of any affine Lie algebra.

The first example of the affine-Sugawara construction 
\cite{BHBardakciHalpern71} used the $su(3)_{1}$ dual quark-model  
currents of the previous section (the dual Coxeter number of $su(n)$ 
is $n$).  The affine-Sugawara construction on the level-one currents 
of affine $so(4,1)$ was also given in our 1971 paper, and extended in the first companion paper \cite{BHHalpern71c} to
the construction on the level-one currents of affine so(3,1)
 and higher Lorentz algebra. The case of $su(n)_{1}$ was given later in \cite{BHHalpern75}. 
We emphasize that the prefactors $(2k+Q_{g})^{-1}$ in these 
 constructions follow from careful consideration of the normal
  ordering \,$:\ldots:$\, of the current
 modes.  The affine-Sugawara constructions, being
quartic in our dual quarks, were called ``current-current'' or
``spin-spin'' interactions in those days, and models which used the dual 
quarks or the affine-Sugawara constructions instead of extra dimensions 
on the open string were known as ``additive'' models.   

In physics the general form (1.11) of the affine-Sugawara construction
was first given by Knizhnik and Zamolodchikov 
\cite{BHKnizhnikZamolodchikov84} thirteen years after our 
examples, and used by these authors to find the general KZ equations.  In 
the same year, the important
paper by Witten  \cite{BHNAB} gave the action formulation and non-Abelian bosonization
of these operator systems, leading to their understanding -- especially 
at higher level -- as interacting
quantum field theories of the WZW type \cite{BHGW}. The independent history of the 
affine-Sugawara constructions in mathematics is noted in Sec. 1.8.

In this connection, we also sketch the history of the $u(n)$ Sugawara-like
constructions in two-dimensional Lorentz-invariant field theories of massless
free fermions: Using a point-splitting regularization of the products of u(n)
currents, Ref. \cite{BHColemanGrossJackiw69} showed the equivalence of the
two-dimensional analogue $\{T_{\mu\nu}\}$ of the original Sugawara
stress-tensor to the free-fermionic stress-tensor \{$\Theta_{\mu\nu}\}$, and
(building on the Abelian case in Ref. Dell'Antonio, Frishman and Zwanziger
\cite{BHDell'AntonioFrishmanZwanziger72}) normal-ordered light-cone analogues
of the u(n) Sugawara construction were later given in
Ref. \cite{BHDashenFrishman73} 
( Eqs. (5.15)-(5.16) of Ref. \cite{BHDashenFrishman75} also gave (and solved) 
light-cone analogues of the $u(n)$ KZ equations for the fermionic 
four-point function.). Again however,
these non-compactified, Lorentz-invariant results had continuous momenta, so
the corresponding affine-Sugawara constructions
do not appear in these papers.  (See also
Refs. \cite{BHHalpern75}, \cite{BHBanksHornNeuberger76}).

 
\section{The Coset Constructions}

The affine-Sugawara construction (1.13) on the currents of affine g is
invariant under Lie g, so we moved next to the study of {\it symmetry
breaking} in current-algebraic CFT on the string.

We begin this discussion with the results in our paper
\cite{BHBardakciHalpern71}. In particular for our example on $su(3)_{1}$, we
first studied the addition to the affine-Sugawara construction of a term $\lambda
J_{8}(m) + \lambda^{2}\delta_{m,0}$ for all constant $\lambda$, which gave
Virasoro generators with {\it continuous} su(3)-symmetry breaking.  This was
the first example of what is today known as inner-automorphic twists or
c-fixed {\it conformal deformations} (Freericks and Halpern
\cite{BHFreericksHalpern88}).

Next we looked for su(3)-symmetry breaking solutions on $su(3)_{1}$ when the
Virasoro generators were assumed to have the more general quadratic form
$L=\sum_{a} C_{a}:J_{a}J_{a}:$ with constant coefficients $\{C_{a},
a=1\ldots8\}$. We stated that solutions beyond the affine-Sugawara
construction for the coefficients $\{C_{a}\}$ existed, giving Virasoro
generators with {\it quantized} $su(3)$-symmetry breaking. This was the first
{\it implicit} mention of coset constructions, but we gave no details about
these solutions -- noting only that they all had ``infinite-degeneracy''
problems. This issue (amplified in our non-additive ``spin-orbit''
constructions of the next section of this reference) was that the solutions to
the Virasoro conditions come in {\it commuting} ``K-conjugate pairs'' of
Virasoro generators $[L(m),\tilde L(n)]=0$ which sum to the affine-Sugawara
construction $L_{g}(m) = L(m)+\tilde L(m)$. This phenomenon was called {\it
K-conjugation covariance} \cite{BHBardakciHalpern71}, \cite{BHHalpern71c},
\cite{BHMandelstam73a}, \cite{BHMandelstam73b}, and the associated spectral
degeneracy of each commuting K-conjugate partner, due to the other, was called
K-degeneracy.
 
 The first {\it explicit} examples of non-Abelian coset constructions were
given in the companion paper ``The Two Faces of a Dual Pion-Quark Model''
\cite{BHHalpern71c}, which solved the Virasoro conditions for all possible (BH
quartic) conformal spin-spin interactions among the four-dimensional dual
quark bilinears T,V,A,P and S (\{$T_{\mu\nu}= :\bar{\psi}
\sigma_{\mu\nu}\psi:\}$ and so on). All 31 of these conformal constructions
were affine-Sugawara or coset constructions, and a number of these were
included explicitly in the applications of the text (see Eqs. (3.7)(3.9) and
(3.14)).  Each of these results was presented as a decomposition of an
affine-Sugawara construction into a commuting K-conjugate pair of conformal
constructions -- which we would today write as 
\begin{equation}
L_{g}(m)= L_{g/h}(m)+ L_{h}(m)\ ,\qquad [L_{g/h}(m),L_{h}(n)]=0\ ,
\end{equation}
where $g$ is the original symmetry algebra, $h$ is any subalgebra of $g$ 
and $g/h$ is the coset space.

We will focus here on the Lie-algebraic identifications in the
simplest case, given as $N^{NS}+N^{5}=N^{H}$ in Eq. (3.14) of the
paper.  All three sets of Virasoro generators are defined in Eqs.
(3.9) and (3.14) as different linear combinations of the BH quartics
$T^{2}$ and $A^{2}$ , and it is not difficult to see that that
$N^{NS}$ and $N^{H}$ are affine-Sugawara constructions respectively on
$so(3,1)_{1}$ and $so(4,1)_{1}$.
Therefore, Eq. (3.14) can be read as the K-conjugate decomposition
 \begin{equation} 
   L_{so(3,1)}(m) + L_{so(4,1)/so(3,1)}(m)\,=\, L_{so(4,1)}(m)
\end{equation}
into the commuting K-conjugate parters on so(3,1) and
so(4,1)/so(3,1). In fact, this result was checked explicitly with the
identities given in the text between the BH dual-quark bilinears T,A
and certain five-dimensional NS bilinears (see Eq. (2.9) of the
text). Using these identities, it was noted that each quartic BH
Virasoro generator collapses to an NS bilinear, the affine-Sugawara
constructions $N^{NS}$ and $N^{H}$ on $so(3,1)_{1}$ and $so(4,1)_{1}$
being equal respectively to the four- and five-dimensional quadratic
NS Virasoro generators, while the non-abelian coset construction
$N^{5}$ (with central charge $c_{so(4,1)/so(3,1)} = 5/2 -2 =1/2$) is
equal to the quadratic Virasoro generators of the fifth NS operator
alone. Similar quartic-to-bilinear equivalences had been seen earlier
in \cite{BHColemanGrossJackiw69}, \cite{BHBardakciHalpern71}.

The general form of the coset construction for all $h\subset g$  
\begin{equation}
L_{g/h}(m)\,=\,L_{g}(m) - L_{h}(m),\,\,\,\,\,\,c_{g/h}\,=\,c_{g} - c_{h}
\end{equation}
was given by Goddard, Kent and Olive \cite{BHGoddardKentOlive85}
fourteen years after our examples. The K-degeneracy of the Virasoro
construction on $g/h$ is now understood as the local gauge invariance
associated to the h-currents of the commuting K-conjugate
affine-Sugawara construction $\{L_{h}(m)\}$ on h . It should be
emphasized however that the coset constructions are by no means the
end of CFT constructions on the currents of affine Lie algebras: Our
original method of solving for the coefficients $\{L^{ab}\}$ in
quadratic forms $\{L^{ab}:J_{a}J_{b}:\}$ eventually led to the
completely K-conjugation-covariant {\it Virasoro master equation}
(Halpern and Kiritsis \cite{BHHalpernKiritsis89}), which includes the
affine-Sugawara constructions and the coset constructions as very
special cases (with rational central charge) of irrational conformal
field theory (Halpern, Kiritsis, Obers and Clubok
\cite{BHHalpernKiritsisObersClubok96}).

\section{The Vertex-Operator Constructions}

There is one more important strand which belongs in this story --
namely the second companion paper in 1975 ``Quantum Solitons which are
SU(N) Fermions'' \cite{BHHalpern75}, which described --using
compactified spatial dimensions on the open string -- the first
vertex-operator constructions of world-sheet fermions and affine Lie
algebra.

By 1972, interest among physicists was shifting back to field theory.
We mention in particular the Bose-Fermi equivalence of the sine-Gordon
and interacting Thirring model studied by Coleman \cite{BHColeman75}
and Mandelstam \cite{BHMandelstam75} in 1975. (See also the free-field
bosonization of the single fermion bilinears by Kogut and Susskind
\cite{BHKogutSusskind75}.) One of us (MBH) realized that -- with two
observations -- this development could be applied to our 1971
non-Abelian current-algebraic constructions on the string.

The first observation was that Mandelstam's normal-ordered
line-integral construction of a single interacting fermion could be
easily generalized to $\it many$ interacting fermions on the line by
including appropriate Klein transformations \cite{BHKlein38}. This led
to bosonic realizations of centrally-extended $su(n)$ current algebras
on the line, and the equivalence of various generalized sine-Gordon
models with corresponding sets of interacting $u(n)$ Thirring
models. In particular, the $su(n)$ currents were constructed from
(n-1) independent two-dimensional bosons, showing the roots of su(n)
in the off-diagonal currents. (See also the extension of these ideas
to the bosonization of two-dimensional non-abelian gauge theories
\cite{BHHalpern76} -- and in particular Eq.(3.13) of this reference.)
 
The second observation in the 1975 paper was that, for massless free
fermions and bosons, there was a structural
parallel between Mandelstam's fermions on the line and particular
string vertex operators on the circle. Indeed, a formal projective map
(from the line to the circle) was given between each two-dimensional
left- and right- mover free-field construction in the field theories
and a {\it pair} of our open-string counterparts on the circle, a
structure which corresponded in fact to closed strings.
These points were emphasized in the second appendix ``Connections with Dual Models'' 
of that paper -- which also gave the vertex-operator 
 construction of many world-sheet fermions
 \cite{BHHalpern75} 
\begin{equation}
\bar{\psi}^{I}(z)\,=\,:\exp\left(i\,Q_{I}(z)\right):\,\bar{\xi}^{I},\,\,\,\,\,
\psi_{I}(z)\,=\,\xi_{I}\,:\exp\left(-i\,Q_{I}(z)\right):.
\end{equation}
where $\{Q_{I}\}$ are the Fubini-Veneziano fields and $\{\bar\xi,
\xi\}$ are the Klein transformations of the text. This relation
guaranteed that the world-sheet fermions were Virasoro primary fields
with conformal weight $1/2$ under the open-string bosonic Virasoro
generators, and we identified these constructions as our
half-integer-moded dual quarks \cite{BHBardakciHalpern71}. 
Although it was not explicitly
mentioned, this identification had in fact been checked
from vacuum expectation values using the natural identification of the
bosonic zero-momentum vacuum $|\{p=J(0)=0\}\rangle$ as the unique
vacuum $|0\rangle$ annihilated by the positive modes of the dual
quarks.

 The vertex-operator construction of $su(n)_{1}$ can therefore
be obtained immediately in a single step by direct substitution of the
fermionic vertex-operators
(1.14) into our dual quark-model construction $\{J_{a}\propto\, :\bar
\psi \lambda_{a} \psi: \}$ of the currents. We regret having omitted
this last simple step in the appendix, thinking it implicit in the
context of the paper: The off-diagonal currents $:\bar
{\psi}^{I}(z)\psi_{J}(z):\,\,\propto\ :exp (i(Q_{I}(z)-Q_{J}(z)):,
I\neq J = 1...n $ of $su(n)_{1}$ follow immediately, showing the roots
of $su(n)$ on (n-1) independent fields, and the diagonal currents of  
$su(n)_{1}$ are a derivative of the same 
(n-1) fields, while the extra $u(1)$ current is a  
derivative of the normalized orthogonal sum $Q_{+}(z) = 
(1/\sqrt{n})\sum_{I}Q_{I}(z)$. We state here the result only for 
$su(2)_{1} + u(1)$  
\begin{eqnarray}
J_{\pm}(z)&=&:\bar{\psi}(z)\,\tau_{\pm}\,\psi(z):\,\, =\,\,:\exp\left(
\pm i\,\sqrt{2}\,Q_{-}(z)\right):,\cr
J_{3}(z)&=&\frac{1}{\sqrt{2}}\,:\bar{\psi}(z)\,\tau_{3}\,\psi(z):\,\,
=\,\,i\,\partial_{z} Q_{-}(z),\cr
J_{0}(z)&=&\frac{1}{\sqrt{2}}\,:\bar{\psi}(z)\,\psi(z):\,\,
=\,\,i\,\partial_{z} Q_{+}(z),\cr
Q_{\pm}(z)&=&\frac{1}{\sqrt{2}}\,\left(Q_{1}(z)\pm Q_{2}(z)\right)
\end{eqnarray}
which is in fact the only case where (as discussed in the text) 
the Klein transformations can be chosen to 
cancel. The charged currents $\{J_{\pm}(z)\}$  of $su(2)_{1}$ show the 
roots of $su(2)$, and results for the neutral current $J_{3}(z)$ and 
the extra $u(1)$ 
current $J_{0}(z)$ are of course closely related to our earlier 
Bose-Fermi relation on the string given in Eq. (1.7) above. 

We concluded that we had  constructed a current-algebraic $su(2)$ from a 
single compactified spatial dimension, and also that
the paper was equivalent to constructing 
as much as current-algebraic $u(22) = su(22) + u(1)$ from the compactified
extra spatial dimensions of the critical open bosonic string. (Quantized
extra momenta were well-known in early open-string theory).
In the following year, Banks, Horn and Neuberger
\cite{BHBanksHornNeuberger76} again studied the bosonization of
$su(n)$ Thirring models, this time with periodic boundary conditions
as $L\rightarrow\infty$ . Translating the fermionic zero modes of
their solution at fixed $L$ into vertex-operators on the string, one
now understands that integer-moded complex versions (CR) of
world-sheet R fermions and $su(n)_{1}$ currents can also be obtained
from the $\it same$ vertex operators by choosing degenerate ground
states with $ \pm1/2$ for each of the $n$ momenta.

In later work (see for example Frenkel, Lepowsky and Meurman
\cite{BHFrenkelLepowskyMeurman88} and Lepowsky \cite{BHLepowsky07}),
one learns that the true vacuum of the full CFT is the BH vacuum
$|0\rangle = |\{p=J(0)=0\}\rangle$ and products of BH currents on the
BH vacuum live on the root-lattice of su(n). The BH theory is itself
reducible, containing many copies of the integrable representations of
su(n) -- each copy living at different $\Delta$'s on the weight
lattice of su(n).

\section{The Confluence with Mathematics}

So ended our three contributions \cite{BHBardakciHalpern71},
\cite{BHHalpern71c}, \cite{BHHalpern75} to current-algebraic internal
symmetry on the string. Loosely speaking, these papers also marked the
end of the first string era, as an increasing number of physicists --
including ourselves -- turned their interest to gauge theory for a decade.

We were therefore fundamentally surprised to see in 1985 that the
seminal paper on the heterotic string (Gross, Harvey, Martinec and
Rohm \cite{BHGrossHarveyMartinecRohm85}) used the 1980 vertex-operator
(root-lattice) construction of $E(8)_{1}$ by Frenkel and Kac
\cite{BHFrenkelKac80}.  Following this, we contacted James Lepowsky
and Igor Frenkel, who patiently exchanged information with us about
these parallel developments in mathematics and physics, including
explanations of the first vertex-operator construction (twisted
$su(2)_{1}$) in mathematics (Lepowsky and Wilson
\cite{BHLepowskyWilson78}), and the earlier Kac-Moody algebras
\cite{BHKac67}, \cite{BHMoody67} -- which include the affine
and hyperbolic algebras as special cases.

In mathematics, our 1971 dual quark-model construction of $su(3)_{1}$
is known as the \emph {first concrete representation of affine Lie algebra},
including the first explicit central term.
In this connection we quote 
first from page 6304 of a 1980 paper by Frenkel \cite{BHFrenkel80}:
``Ten years ago two physicists, Bardakci and Halpern, in
Ref. \cite{BHBardakciHalpern71} constructed a representation of the
subalgebra $\tilde{g}l(l)$ of $\tilde{o}(2l)$ in the space
$V((2\mathbb{Z}+1)^{l}))$ (see formulas 3.1-3.11 of
Ref. \cite{BHBardakciHalpern71}). At that time the theory of affine
Lie algebras began to take its first steps.'' A less technical form of this
statement is found on page 36 of the Introduction to the review article
by Frenkel, Lepowsky and Meurman \cite{BHFrenkelLepowskyMeurman88}:
``In their proposal of current-algebraic internal symmetry for strings, 
Bardakci and Halpern \cite{BHBardakciHalpern71}, independently of 
mathematicians, discovered affine Lie
algebras in 1971, including the first case of a concrete representation
--- a fermionic realization of $\hat{sl(3)}$; irreducibility issues
arose later in independent discoveries by mathematicians.''

The affine-Sugawara constructions (\cite{BHBardakciHalpern71}, 
\cite{BHHalpern71c}, \cite{BHKnizhnikZamolodchikov84}) 
in physics were also found in math \cite{BHSegal81}, 
where they are known as the Segal operators.

The early vertex-operator constructions in physics are also well-known 
in mathematics. In this connection, we find
again on the same page of Ref. \cite{BHFrenkelLepowskyMeurman88}: 
``The untwisted vertex-operator construction
of $\hat{sl(n)}$-representations from compactified spatial dimensions
was implicit in \cite{BHHalpern75}, \cite{BHBanksHornNeuberger76}.''
A supplementary statement is found on page 365 of 
Ref. \cite{BHLepowsky07}: ``It turned out that vertex operators and
symmetry are closely related. Instances of this had already been
discovered in physics, including the works \cite{BHHalpern75} and
\cite{BHBanksHornNeuberger76}.''.

In conclusion, one might say that for us, ignorance was bliss -- or,
more optimistically -- that physicists can create mathematics. In
either case, it is clear that both fields have profited handsomely from this
confluence.

\section{The Larger Perspective} 

In our historical discussion above we have focused on the discovery of current-algebraic CFT on the string
worldsheet, but to complete the modern perspective we should  
emphasize that this is just one of many mathematical methods which have been found
and developed in string theory. Over the last 25 years, this domain has been
the main source of new methods in theoretical physics that deeply changed the discipline
and had many other applications in different areas, such as statistical mechanics
and condensed matter physics. These developments involved the latest mathematical 
advances in algebra, topology and geometry.
Some of these results (such as the affine algebras described here) were first discovered in string theory
and then studied systematically by mathematicians, while others were imported from mathematics and made
clear and practical by the string applications.

This communication and cross-fertilization has also had a large impact 
in mathematics, where the field theory methods are now well appreciated
(see for example the lecture notes in Ref. \cite{BHPrincetonLect}).
In the 80's and 90's, Edward Witten has played a crucial role in this 
interdisciplinary endeavor.

In our context we note in particular the important set of string-theory mathematical
methods given by general two-dimensional conformal field theory,
which began with the analysis by Belavin, Polyakov and
Zamolodchikov \cite{BHBPZ}. They used the mathematical
results of Kac \cite{BHKac90} and Feigin and Fuchs \cite{BHFeiginFuchs}
on representation theory of the Virasoro algebra, and realized their
relevance for exact descriptions of two-dimensional critical phenomena
(see for example the collection of important papers in \cite{BHItzyksonSaleurZuber88}).

For help in the preparation of this article, we thank O. Ganor, I.
Frenkel, Y. Frishman and J. Lepowsky.

\end{document}